\documentstyle[twocolumn,aps]{revtex}
\begin{document}
\draft

\title{Large-basis shell model studies of light nuclei with a
multi-valued $G$-matrix effective interaction}

\author{D. C. Zheng and B. R. Barrett}
\address{Department of Physics, University of Arizona,
	Tucson, Arizona 85721}

\author{J. P. Vary}
\address{Department of Physics and Astronomy, Iowa State University,
	Ames, Iowa 50011}

\author{W. C. Haxton and C.-L. Song}
\address{Institute for Nuclear Theory and Department of Physics,
	University of Washington, Seattle, Washington 98195}

\date{\today}
\maketitle

\begin{abstract}
Large-basis shell model studies of low-lying
excitations in light nuclei from $^4\mbox{He}$ to $^7\mbox{Li}$ have been
performed with a multi-valued $G$-matrix effective interaction,
as recently suggested by Haxton {\it et al.}.
Calculations were performed relative to the vacuum (``no core") using
very large, separable model spaces containing all excitations with
unperturbed energies up to $8\hbar\Omega$.  Using $G$ matrices
derived from a new Nijmegen potential,
we achieve a very satisfactory description of these excitations.
\end{abstract}
\pacs{21.60.Cs, 21.10.Ky, 27.10.+h}

\section{Introduction}
The nuclear Hamiltonian
\begin{equation}
H = \frac{1}{2} \sum_{i\not=j}^A \left( T_{ij} + V_{ij} \right), \label{SMH}
\end{equation}
where $T_{ij}$ is the relative kinetic energy operator and $V_{ij}$ the
nucleon-nucleon ($NN$) interaction, is often treated in the nuclear shell
model by introducing the one-body harmonic oscillator (HO) Hamiltonian
\begin{equation}
H_0 = \sum_{i=1}^A h_i = \sum_{i=1}^A \left(\frac{\bbox{p}_i^2}{2M} +
\frac{1}{2}M\Omega^2\bbox{r}_i^2 \right)   \label{H0}
\end{equation}
to classify the many-body states: Slater determinants are formed
from the products of these single-particle wave functions.  These
many-body basis states can be labeled according to the number of
oscillator quanta they contain, $N_{\rm sum} = \sum_{i=1}^A N_i$,
or, equivalently, the unperturbed energies
\begin{equation}
\sum_{i=1}^A \left(N_i + \frac{3}{2}\right) \hbar\Omega ,
\end{equation}
where $N_i$ is the number of oscillator quanta ($2n_i+l_i$) of the $i$th
single-particle state.  Conventionally the labeling is relative to the minimum
energy Slater determinant(s), so that the basis states are
partitioned into $0\hbar\Omega$, $1\hbar\Omega$, $2\hbar\Omega$, etc.,
configurations.

Early shell model calculations were generally restricted to a
single shell, such as the $0p$ or $1s\,0d$ shells, and thus involved
only $0\hbar\Omega$ valence nucleon configurations.  An effective
interaction is then introduced to account for the effects of
excluded configurations, including very high energy excitations
associated with the hard core in the $NN$ interaction.  The
lowest order approximation to this effective interaction is the
two-body $G$ matrix, which links two-particle states within the
model space by a ladder series for scattering in the excluded
space.  The resulting interaction $V^{\rm eff}(ab;cd)$ is a
function of the valence-shell single particle states ${c,d}$
and ${a,b}$ that label the starting and ending states of
the ladder, respectively.

In recent years shell model calculations involving two or more
major shells have been frequently performed.
A full multi-$\hbar\Omega$ basis is one that includes all many-body
configurations, such that $N_{\rm sum} \leq N_{\rm {max}}$ for some
$N_{\rm {max}}$. For example, a calculation of the positive-parity states
in $^{16}\mbox{O}$ might
include all $(0+2)\hbar\Omega$ or $(0+2+4)\hbar\Omega$
many-body configurations, relative to the closed core (fully
occupied $0s$ and $0p$ shells). Because of the importance of
the nuclear mean field, such a truncation provides a reasonable
starting point for describing the ``long-wavelength" properties
of nuclei.  A $(0+2+4)\hbar\Omega$ calculation (for which $N_{\rm max} = 16$)
of $^{16}\mbox{O}$ yields a reasonable description of low-lying
excitations, including effects associated with highly deformed
excited states \cite{hj}.

Such full multi-$\hbar\Omega$ bases have other appealing properties.  If
HO single-particle states are employed, the model space
wave functions can be decomposed so that the relative degrees
of freedom are separated from a pure oscillator state
center-of-mass component.  Thus the overcompleteness of the
Slater determinants (which depend on 3$A$ coordinates, while
intrinsic wave functions depend on 3$(A-1)$)
can be cured by retaining only those linear
combinations which keep the center of mass in the $0s$ state.

A second property has to do with technical difficulties in
evaluating the effective interaction.  If, in addition to defining the
basis states, $H_0$ of Eq.~(\ref{H0}) plays the role of the unperturbed
Hamiltonian, then the unperturbed energies of configurations
in the excluded space always exceed those in the model space.
In the case of parity-conserving (parity-nonconserving)
interactions, the minimum gap is $2\hbar\Omega$ ($1\hbar\Omega$).
This contrasts sharply with other choices for the model space.
For example, a partial $1\hbar\Omega$ calculation of the negative
parity states in $^{12}\mbox{C}$ in which the valence nucleons are restricted
to the $0p$ and $1s\,0d$ shells leads to intermediate states in the
core-polarization process (bubble diagram) with vanishing energy denominators:
particle-hole excitations in the excluded space of the form
$0p(0s)^{-1}$ have the same energy, $1\hbar\Omega$, as those
in the model space.  While one might attempt to cure this problem
by introducing a spin-orbit interaction in $H_0$ to break the
degeneracy, this tends to produce small energy denominators of
somewhat random sign, leading to serious convergence problems.
Thus one sees that the gaps characterizing complete multi-$\hbar\Omega$
bases are quite attractive.  As discussed in Ref.~\cite{haxton}, this
nice feature can be perserved order by order in calculations
of the full $V^{\rm eff}$, provided a suitable perturbation scheme
for $V^{\rm eff}$ is employed.

Investigators doing large-basis shell model calculations
in multi-$\hbar\Omega$ spaces have consistently chosen
effective two-body interactions of the form $V^{\rm eff}(ab;cd)$,
just as in traditional $0\hbar\Omega$ calculations.
The appropriate effective
interaction \cite{haxton} in such spaces must carry an additional index
$N^{\rm spectators}_{\rm sum}$
$V^{\rm eff}(ab;cd;N^{\rm spectators}_{\rm sum})$,
where $N^{\rm spectators}_{\rm sum}$ labels the total
oscillator quanta of the ``spectator'' (i.e., non-interacting) nucleons
in the many-body states connected by the matrix element $V^{\rm eff}(ab;cd)$.
It is given by
\begin{equation}
N^{\rm spectators}_{\rm sum}
= N_{\rm sum} - N_c - N_d = N'_{\rm sum} - N_a - N_b ,
\end{equation}
where $N_{\rm sum}$ and $N'_{\rm sum}$ are the numbers of
the total oscillator quanta of the initial and final many-body states,
respectively. In the case of traditional $0\hbar\Omega$ calculations, all
basis states are characterized by the same $N^{\rm spectators}_{\rm sum}$,
so this additional index is unnecessary.  But for multi-$\hbar\Omega$ bases,
the $N^{\rm spectators}_{\rm sum}$ dependence is essential:
if model-space configurations
exist with different unperturbed energies, the gaps and interactions
coupling these configurations to the excluded space will differ.
The appropriate energy denominators in the $G$-matrix ladder sum
are not given just by the initial and final two-particle labels,
but also depend on the energies of the $A-2$ ``spectator" nucleons.

The omission of the $N^{\rm spectators}_{\rm sum}$ dependence in
recently reported large-space shell-model calculations \cite{lighta,mfd}
amounts to neglect of certain many-body processes of the same
unperturbed energy as some retained many-body processes. While
the effects of these neglected many-body processes are expected
to decrease in importance as the number of shells included in the model
space increases, our investigation here, which retains them through the
$N^{\rm spectators}_{\rm sum}$ dependence
of the two-body effective interaction, will
reveal that these neglected effects are important in present-day
calculations.

We shall see in the following calculations that the resulting
shifts can be large, amounting to about 5 MeV for diagonal matrix
elements.  The approximation in present-day multi-shell calculations
to neglect the $N^{\rm spectators}_{\rm sum}$
dependence can lead to unattractive
consequences.  One example is the apparent need for unrealistic single
particle energies to reduce the splittings between the
$0\hbar\Omega$ and $2\hbar\Omega$ states, as required by experiment.

In this paper we present the results of multi-$\hbar\Omega$
shell model calculations for $^4\mbox{He}$, $^5\mbox{He}$, $^6\mbox{Li}$,
and $^7\mbox{Li}$ in which the two-body effective interaction is evaluated
with full $N^{\rm spectators}_{\rm sum}$ dependence.
As a result, we obtain a lowering of states that are
dominated by $1\hbar\Omega$ and $2\hbar\Omega$ configurations,
relative to $0\hbar\Omega$ states.  This improves the
agreement with experiment.  The calculation of the $G$ matrix
is described in Sec.~II.  This work is distinguished from our
previous studies \cite{lighta,mfd} in another important aspect, namely,
the extension of the model spaces for light nuclei to include
excitations up to $8\hbar\Omega$.  These calculations are
``no core," performed relative to vacuum and, of course,
include excitations out of the $0s$ shell.  The results
are presented in Sec.~III, where a comparison with previous
calculations is also made.  The dependence of the results
on the size of the model space is discussed in Sec.~IV,
and the consequences of neglecting the
$N^{\rm spectators}_{\rm sum}$ dependence of
the $G$ matrix explored.  Our conclusions are given in Sec.~V.

\section{Multi-valued $G$-Matrix Effective Interactions}
\label{mG}

Shell model diagonalizations of the Hamiltonian in Eq.~(\ref{SMH}) are
performed within truncated Hilbert spaces containing, hopefully,
most of the long-wavelength modes important to describing
properties such as nuclear sizes, low-lying excitations
and collective modes.  The neglected degrees of freedom, e.g.,
those high-momentum interactions arising from $NN$ interactions
at short distances, must be incorporated into the calculation
through effective interactions (and effective operators).
While in principle an effective interaction exists that will
reproduce exact eigenvalues in a model space calculation,
in practice it can only be evaluated approximately.

The shell model Hamiltonian we diagonalize is
\begin{eqnarray}
H_{\rm {SM}} &=& \frac{1}{2} \sum_{i\not=j}^A \left(T_{ij}
	+ V_{ij}^{\rm eff}(N^{\rm spectators}_{\rm sum}) \right)
	+ V_{\rm {Coulomb}} \nonumber \\
& & + \lambda \left( H_{\rm {c.m.}}-\frac{3}{2}\hbar\Omega \right),
					\label{hsm}
\end{eqnarray}
where $T_{ij} = \frac{1}{2AM} (\bbox{p}_i-\bbox{p}_j)^2$ and
\begin{equation}
\frac{1}{2} \sum_{i\not=j}^A T_{ij}
	= \sum_{i=1}^A \frac{\bbox{p}_i^2}{2M} - T_{\rm {c.m.}}
\end{equation}
with $T_{\rm {c.m.}}=\frac{1}{2AM} \left(\sum_{i=1}^A \bbox{p}_i \right)^2$.
Note that $H_{\rm {c.m.}} = T_{\rm {c.m.}}+U_{\rm {c.m.}}$ where
$U_{\rm {c.m.}}=\frac{AM\Omega^2}{2}\left(\sum_{i=1}^A \bbox{r}_i\right)^2$.
The last term in Eq.~(\ref{hsm}) is
included in order to project out spurious center-of-mass
motion: inclusion of this term with a large value of $\lambda$
produces low-lying excitations with the center-of-mass in the
$0s$ state.  For this procedure to work properly, the model
space must be exactly separable, as is the case for complete
multi-$\hbar\Omega$ bases.

The bare Coulomb interaction $V_{\rm {Coulomb}}$ is diagonalized only
within the model space.  For the strong potential, the
effective interaction evaluated at the two-body level has
the familiar Bruckner $G$-matrix \cite{bruckner} form, but with an important
difference in the definition of the Pauli exclusion operator $Q$,
\begin{eqnarray}
& & G(E_0,N^{\rm spectators}_{\rm sum}) =
 V_{12} + \tilde{V}_{12} Q(N^{\rm spectators}_{\rm sum}) \nonumber \\
& & \;\;\;\; \times \frac{1}{E_0 - (h_1 + h_2 + \tilde{V}_{12})}
	Q(N^{\rm spectators}_{\rm sum}) \tilde{V}_{12}, \label{G}
\end{eqnarray}
where $E_0$ is the energy of the initial two-body
state (i.e., the starting energy), $V_{12}$ is the bare $NN$ potential,
$\tilde{V}_{12} = V_{12} - U_{12}$, $U_{12} = \frac{M\Omega^2}{2A}
({\bf r_1}-{\bf r_2})^2$, and $Q(N^{\rm spectators}_{\rm sum})$ is the Pauli
operator that restricts all intermediate states to lie in
the Pauli-allowed, excluded space.  In conventional treatments of the Brueckner
$G$ matrix, a very similar equation arises, but with a $Q$
that excludes certain intermediate states based only on the
single-particle labels of the two particles involved in the
scattering. There is no $N^{\rm spectators}_{\rm sum}$ dependence in $Q$ for
single shell calculations but, as discussed above, this
$N^{\rm spectators}_{\rm sum}$ dependence arises in multi-$\hbar\Omega$
calculations.

The index $N^{\rm spectators}_{\rm sum}$ signifies the role of the full
model space many-body configuration in controlling the
intermediate two-particle states available for scattering.
In a shell model calculation whose model space includes all many-body states
with $N_{\rm sum} \leq N_{\rm max}$, the allowed intermediate states for
the two particles, ``1'' and ``2'', scattered by $\tilde{V}_{12}$,
are specified by:
\begin{equation}
N_1+N_2 + N^{\rm spectators}_{\rm sum} > N_{\rm {max}}\; ,
\end{equation}
which corresponds to the following Pauli operator:
\begin{equation}
Q(N^{\rm spectators}_{\rm sum}) = \left\{ \begin{array}{l}
	0 \;\; {\rm if} \;\; N_1+N_2
	\leq N_{\rm {max}} - N^{\rm spectators}_{\rm sum} , \\
	1 \;\; {\rm otherwise.} \end{array} \right. \label{Q0A}
\end{equation}
In Fig.1, we depict the various spectator-dependent Pauli operators
appropriate for a full $6\hbar\Omega$ calculation of $^6\mbox{Li}$
($N_{\rm max}=8$).

The fact that we introduce a spectator dependence to the $G$-matrix raises
interesting possibilities for identifying specific Pauli-violating processes.
Some two-particle scattering states in the excluded space will place a nucleon
in a single-particle state that may be occupied by a spectator nucleon in a
given model space wavefunction.  One might avoid these Pauli violating
processes in a full multi-$\hbar\Omega$ calculation by labeling $Q$
with the full
set of quantum numbers on which $G$ operates.  This, of course, is impractical.
However, for the specific case of these light nuclei and for
$N^{\rm spectators}_{\rm sum} = 0$,
we can easily eliminate the Pauli violating processes involving the $0s$
nucleons by including the ``wings'' as depicted in Fig.~1.  However, we have
found that the presence or absence of the wings in the case
$N^{\rm spectators}_{\rm sum} = 0$ results in minor differences in our results
due to the large size of the model spaces.

To provide the reader with some measure of the size of the effects associated
with $N^{\rm spectators}_{\rm sum}$, we give in Table I the matrix elements
$\langle (0s_{1/2}\,0s_{1/2}) | V^{\rm eff} | (0s_{1/2}\,0s_{1/2}) \rangle$,
$\langle (0s_{1/2}\,0p_{3/2}) | V^{\rm eff} | (0s_{1/2}\,0p_{3/2}) \rangle$,
and
$\langle (0p_{3/2}\,0p_{3/2}) | V^{\rm eff} | (0p_{3/2}\,0p_{3/2}) \rangle$,
that we evaluated for a full $6\hbar\Omega$ calculation of the
positive-parity states in $^6\mbox{Li}$
(which we will discuss in Sec.~\ref{Li6s}). In this calculation,
$N_{\rm max} = 8$ and $N_{\rm sum}$ can take on four values (2, 4, 6, 8).
The Table shows that the values of these diagonal matrix elements can shift
by up to 3.3 MeV when $N^{\rm spectators}_{\rm sum}$ dependence is
properly treated.

This ``multi-valuedness" is a bookkeeping complication in
shell model studies.  However, its inclusion builds in
essential physics previously missing from multi-$\hbar\Omega$
calculations.  Model-space states of higher unperturbed energy
are now more strongly repelled downwards by effects
of states in the excluded space which are included in $G$ for the first time.
For example, in a (0+2+4)$\hbar\Omega$ calculation of $^{16}\mbox{O}$,
the contribution to $G$ that is second order in $\tilde{V}_{12}$
contributes to shifts in the position of the 0$\hbar\Omega$
configuration only because of potential matrix elements
with an unperturbed energy denominator of
$6\hbar\Omega$ or larger. However, the 4$\hbar\Omega$
configurations are shifted by matrix elements with an energy
denominator of $2\hbar\Omega$ or larger.
The larger shifts result from a more complete inclusion of intermediate
two-particle states scattering via a spectator dependent definition
of $Q$, which reduces the size of
the $Q$=0 region as $N^{\rm spectators}_{\rm sum}$ increases.  Since $G$
for a larger $Q$=0 region is less ``attractive'', numerically, we find
that states that are predominantly of $1\hbar\Omega$ and
2$\hbar\Omega$ character, which were obtained at energies
a few MeV too high in Ref.~\cite{mfd}, are now lowered
relative to states that are predominantly $0\hbar\Omega$ in
character.

The intent of the present study is to illustrate the effects
attributable to the multivaluedness of $G$.  We therefore
follow the treatment in Ref.~\cite{mfd} in other respects,
which includes two approximations in the evaluation of
Eq.~(\ref{G}).  The first is the substitution of $V$ for $\tilde{V} = V-U$.
This considerably improves the convergence of
the numerical procedures we employ in evaluating $G$, since
the growth of the HO $U(r)$ at large $r$ is troublesome.
Earlier studies \cite{jaqua} have shown that neglecting $U$
in terms second order or higher in $V$ induces errors in
calculated binding energies of a few hundred keV; these errors
decrease as the size of the model space increases.

The second is the replacement of the ``starting energy'' $E_0$
in the ladder sum of Eq.~(\ref{G}) by
\begin{equation}
\omega_{cd} = \epsilon_c + \epsilon_d + \Delta \, ,
\end{equation}
where $(\epsilon_c+\epsilon_d)$ is the unperturbed energy of the
initial two-body state and $\Delta$ is a parameter whose value is
adjusted to yield a reasonable binding energy. This substitution was
introduced and explained in Ref.~\cite{mfd}:
it is a phenomenological correction for the omission of folded diagrams
and higher-order contributions to the effective interaction, those beyond $G$
that involve the multiple scattering of clusters of three or
more nucleons.  As some of these omitted corrections will shift
starting energies from their unperturbed values towards the
true eigenvalues, it is not surprising that
a phenomenological shift $\Delta$ is necessary to reproduce
experimental binding energies when a two-body $G$ matrix is used.
However, energy differences are relatively insensitive to the choice of
$\Delta$. Such a state-dependent choice for the starting energy leads to a
non-Hermitian $G$ matrix. But the non-Hermiticity is found to be
small and we obtain a Hermitian effective interaction
by symmetrizing the $G$ matrix:
\begin{eqnarray}
& & \langle ab |V^{\rm eff}(N^{\rm spectators}_{\rm sum})| cd \rangle_{J,T}
						\nonumber \\
&=& \frac{1}{2}\left[ \langle ab |G(\omega_{cd},N^{\rm spectators}_{\rm sum})|
	 cd \rangle_{J,T}\right.          \nonumber \\
&+& \left. \langle ab |G(\omega_{ab},N^{\rm spectators}_{\rm sum})
	| cd \rangle_{J,T} \right].	   \label{GGG}
\end{eqnarray}

We employ the method of Barrett {\it et al.} \cite{bhm}
to calculate the $G$ matrices.
For the bare $NN$ interaction $V_{12}$ in Eq.~(\ref{G}),
we use the non-relativistic version
of the new Nijmegen potential (NijmII) \cite{nijmii}. The HO basis
parameter $\hbar\Omega$ is fixed at 14 MeV. Calculations for a different
choice of $\hbar\Omega$ are performed for selected cases for the
purpose of comparison.

\section{Results and Discussion}
\label{results}
The shell-model calculations are performed for $^4\mbox{He}$,
$^5\mbox{He}$, $^6\mbox{Li}$, and $^7\mbox{Li}$
in large, no-core, model spaces using the Many-Fermion-Dynamics
Shell-Model code \cite{code}. We properly evaluate the $G$ matrix
for full multi-$\hbar\Omega$ spaces, resulting in a multi-valued two-body
effective interaction.  The calculated results
are presented in Tables I to IV which we discuss below.

\subsection{$^4\mbox{He}$}  \label{he4}
For the positive-parity states in $^4\mbox{He}$, we use a 9-major-shell
model space which allows us to include all configurations
with $N_{\rm sum}=N_1+N_2+N_3+N_4 \leq 8$ [i.e., $N_{\rm {max}}$ = 8].
For the negative-parity states, we use a 8-major-shell
space and include all configurations with
$N_{\rm sum} \leq N_{\rm {max}} = 7$.
The lowest configuration in this nucleus is $(0s)^4$, which has
$N_{\rm sum}=0$,
so we are doing a full $8\hbar\Omega$ ($7\hbar\Omega$)
calculation for the positive-parity (negative-parity) states.
The calculations involve $(N_{\rm {max}}+1)$
$G$ matrices, corresponding to $(N_{\rm {max}}+1)$ possible values
of $N^{\rm spectators}_{\rm sum}$ (from 0 to $N_{\rm {max}}$).

The parameter $\Delta$ in the starting energy is chosen to be -55 MeV,
which yields a reasonable binding energy of 26.3 MeV.
(It should be pointed out that due to the large size of
the model space, the $G$ matrix elements are a very smooth function of
$\Delta$. The binding energy of $^4\mbox{He}$ increases by less than
1 MeV when $\Delta$ is increased by 10 MeV from -60 MeV to -50 MeV.
See Ref.~\cite{mfd} for a discussion of the sensitivity of the
results to $\Delta$.) The calculated results are given in Table II
and plotted in Fig.2 along with the experimental data,
taken from a recent compilation of Tilley {\it et al.} \cite{he4exp}
and Ref.\cite{rms1}.

As can be seen from Table II and Fig.2, very good agreement with
experiment is obtained for the energy spectrum. In particular,
the experimental low-lying negative-parity (``$1\hbar\Omega$'') states
are reproduced to within 1.2 MeV with
a correct level sequence. The first excited $0^+$
(predominantly ``$2\hbar\Omega$'') state
is obtained at an excitation energy
of 21.8 MeV, only 1.6 MeV higher than experiment.

The importance of the high-energy configurations can be seen by
examining the wave functions. In terms of major-shell
configurations, the calculated g.s.~wave function can be expressed as
\begin{eqnarray}
[0^+_1] & \simeq & 70\% |0\hbar\Omega\rangle
                + 14\% |2\hbar\Omega\rangle
                +  9\% |4\hbar\Omega\rangle \nonumber \\
            & & +  3\% |6\hbar\Omega\rangle
                +  4\% |8\hbar\Omega\rangle,  \label{gs14}
\end{eqnarray}
while for the first excited state, we obtain
\begin{eqnarray}
[0^+_2] & \simeq & 8\% |0\hbar\Omega\rangle
                +  61\% |2\hbar\Omega\rangle
                +  15\% |4\hbar\Omega\rangle \nonumber \\
            & & +  13\% |6\hbar\Omega\rangle
                +   3\% |8\hbar\Omega\rangle.  \label{es14}
\end{eqnarray}
As can be expected, the $0^+_1$ state is dominated by the
$0\hbar\Omega$ configuration and the $0^+_2$ state is dominated
by the $2\hbar\Omega$ configuration. However,
we see from the above ``wave functions'' that
$|0^+_1\rangle$ has significant
$2\hbar\Omega$ and $4\hbar\Omega$ admixtures
while $|0^+_2\rangle$ has significant
$4\hbar\Omega$ and $6\hbar\Omega$ admixtures.
Therefore, for a reasonable description of the $0^+_1$ and $0^+_2$
states {\em using a HO basis} with $\hbar\Omega=14 {\rm {MeV}}$,
one needs to perform a
$4\hbar\Omega$ calculation and a $6\hbar\Omega$ calculation, respectively.
The requirement of a large HO space for convergence of wave functions
dominated by $1\hbar\Omega$ and $2\hbar\Omega$ components has been
established by Ceuleneer {\it et al.} in Ref.~\cite{ceul}
where a $10\hbar\Omega$ calculation was performed
for $^4\mbox{He}$ using a modified Sussex \cite{sussex} interaction.
Because the HO potential is too confining at large distances,
high-lying configurations are required to properly describe the
shape of the nuclear surface. Alternatively, one may address these same
physics issues within the effective
Hamiltonian formalism in a HO space by evaluating the contributions
of effective many-body interactions.

The weights of the different major-shell
configurations listed in Eq.~(\ref{gs14}) and Eq.~(\ref{es14}) depend on
the choice of single-particle basis: they would change if we
were to adopt a Hartree-Fock basis or, even, retain HO wave
functions but change the value of the oscillator parameter.
One procedure for removing this arbitrariness, for a given model
space, would be to diagonalize the ground state one-body
density matrix, then transform to a new basis given by the
eigenvectors.  The naive ``closed shell'' would then be defined
by the largest eigenvalues.  We will discuss these issues
further in Sec.~\ref{size}.

The major differences in the spectra resulting from the present work and
Refs.\cite{lighta,mfd} appear in the lowering of the excited states
due to increased admixtures of higher lying configurations for the reasons
mentioned above. For example, the $0_2^+$ state is lowered by
11.9 MeV from its excitation energy in Ref.\cite{lighta}
and by 4.3 MeV from its excitation energy in Ref.\cite{mfd}.  On the
other hand, the $0_1^-$ state is lowered by only 0.8 MeV and 1.3 MeV
relative to its excitation energy in Ref.\cite{lighta} and
Ref.\cite{mfd}, respectively.

Since there are a number of differences between the present work and our
previous efforts, we will discuss in Sec.~\ref{size}
the dependence of our results on model
space size alone with all other ingredients in the calculations held fixed.

\subsection{$^5\mbox{He}$}
Throughout this work, the unperturbed energy of a configuration is measured
with respect to that of the lowest configuration (of {\em either} parity).
For $^5\mbox{He}$, the lowest ($0\hbar\Omega$) configuration is
$(0s)^4(0p)^1$ with $N_{\rm sum}=1$.
We do a full $6\hbar\Omega$ calculation
($N_{\rm {max}}=7$) for the negative-parity states and a
full $7\hbar\Omega$ calculation ($N_{\rm {max}}=8$)
for the positive-parity states. The parameter $\Delta$
in the starting energy is taken to be -45 MeV for this and
other $0p$-shell nuclei considered in this work.
The results are shown in Table III and
in Fig.3. The first excited state ($1/2^-$)
is obtained at an excitation energy of 2.47 MeV.
The well-known $3/2^+$ state at 16.75 MeV is reproduced
at an energy of 19.06 MeV. We have also obtained a number of low-lying
``$1\hbar\Omega$'' positive-parity and ``$2\hbar\Omega$''
negative-parity states that have not yet been identified
experimentally. There is a $1/2^+$ state at an excitation energy of
only 4.34 MeV and there are two nearly degenerate $3/2^+$ and $5/2^+$
states at about 9.7 MeV.
The lowest ``$2\hbar\Omega$'' state ($3/2^-$) is obtained at
an excitation energy of 12.01 MeV.

The energy splitting $\Delta E$ between the $1/2^-$ state
and the $3/2^-$ state (g.s.) is of particular interest.
Analyses of experimental data yielded many different values for
$\Delta E$, ranging from 1.4 MeV \cite{arndt} to more than 5 MeV \cite{woods}.
In a recent Green's function Monte Carlo (GFMC) calculation \cite{GFMC}
where the odd neutron is restricted to be in the $p_{1/2}$ or
$p_{3/2}$ state (not pure HO $0p$ states),
a small splitting of 0.8 MeV is obtained. In our
calculations, we notice that $\Delta E$ tends to decrease
as we include more $p$ orbitals in the model space.
The values for $\Delta E$ obtained in the
$0\hbar\Omega$, $2\hbar\Omega$, $4\hbar\Omega$ and $6\hbar\Omega$
calculations are 2.81, 3.15, 2.89 and 2.47 MeV, respectively.
As this series does not appear to have converged, it is quite
possible that still larger model spaces would yield a result
below the $6\hbar\Omega$ value of 2.47 MeV.

Our predictions of the low-lying positive-parity states agree quite well
with other theoretical works \cite{hees,barker} where a $1/2^+$ state
at about 5-7 MeV is predicted along with two additional states
($3/2^+$ and $5/2^+$) at about 12-14 MeV, except that our results
are slightly lower. These levels were first obtained by van Hees and
Glaudemans \cite{hees} in a $(0+1)\hbar\Omega$ shell-model calculation using
a phenomenological interaction
(obtained by fitting selected nuclear properties)
and were later supported by other shell-model
calculations using different phenomenological interactions \cite{barker}.
These states are expected to be broad and cannot be easily identified
experimentally. However, they can be seen in an R-matrix analysis
of the nucleon-alpha phase shifts with a channel radius of
$a \sim 5\, {\rm {fm}}$ \cite{barker}, but not with a smaller $a$ of
about 3 fm commonly used before. A large channel radius of
$5.5 \pm 1.0\, {\rm {fm}}$ has been determined from the stripping and
pickup reaction data \cite{woods}.

These low-lying positive-parity states were obtained at higher energies in
our previous $3\hbar\Omega$ (i.e., $N_{\rm {max}} = 4$) single-$G$ calculation
\cite{mfd}. This can be explained by noting that the calculated wave
functions of these states contain significant higher-shell configurations.
For example, for the $1/2^+$ state, we obtain
\begin{equation}
[1/2^+_1] = 45\% |1\hbar\Omega\rangle + 28\% |3\hbar\Omega\rangle
		 +18\% |5\hbar\Omega\rangle +  9\% |7\hbar\Omega\rangle \; .
				\label{half}
\end{equation}
The calculated wave functions also show that these states can be
roughly described as systems with one neutron moving in an $s$ or $d$
(not necessarily $1s$ or $0d$) orbitals outside a $^4\mbox{He}$ core.

The point made by Eq.~(\ref{half}) and the associated discussion may appear
provocative. If we try to interpret all our states as predictions for the
locations of resonances, this would indicate the near absence of a shell gap
in $^5\mbox{He}$.  However, we should always keep in mind that
$^5\mbox{He}$ is unbound with
respect to neutron emission and that all the states of
$^5\mbox{He}$ are experimentally above breakup threshold
(i.e. in the continuum).
Thus, as we systematically expand our model space we would expect our
calculated results to approach a continuous spectrum,
though presumably the convergence
might be quite slow due to the use of confined HO
wave functions in the shell model expansion. By analyzing transition
strength functions we would be able to isolate those states which are truly
predicted resonances from the background of continuum states.  However, at the
present time, our model space is too limited to be able to carry out such an
analysis.  Nevertheless, we expect that states which are experimentally narrow
will be reproduced by our theoretical framework.

The calculated $3/2^+$ state at 19.06 MeV is dominated by the configuration
$(0s)^3 (0p)^2$, which is basically the ground state of $^6\mbox{Li}$
with a proton removed from the $0s$ orbital. It can therefore be
identified as the 16.75 MeV state observed experimentally in
nucleon knock-out reactions with a $^6\mbox{Li}$ target \cite{a5to10}.

The lowest ``$2\hbar\Omega$'' state is calculated at a surprisingly
low excitation energy of 12.01 MeV with $(J^{\pi},\; T) = (3/2^-,\; 1/2)$.
This state was obtained at a much higher energy of 21.5 MeV
in a $4\hbar\Omega$ (i.e., $N_{\rm {max}} = 5$) single-$G$
calculation \cite{mfd}.
The dramatic decrease of the excitation energy is due, again, to
the importance of $6\hbar\Omega$ admixtures,
as can be seen from the following decomposition:
\begin{equation}
\left[ 3/2^-_2\right] = 1\% |0\hbar\Omega\rangle + 52\% |2\hbar\Omega\rangle
		+28\% |4\hbar\Omega\rangle + 19\% |6\hbar\Omega\rangle \; .
\end{equation}
The fact that the above wave function contains a large $6\hbar\Omega$ component
implies that the energy of this state, now at 12.01 MeV, is likely
to be further decreased if one is able to do an even larger calculation to
include the $8\hbar\Omega$ ($N_{\rm max}=9$) configuration.
The second lowest ``$2\hbar\Omega$'' state is obtained at
15.21 MeV with $(J^{\pi},\; T) = (1/2^-,\; 1/2)$.
These two ``$2\hbar\Omega$'' states were also obtained by
Wolters {\it et al.} \cite{wolters}
in a $(0+2)\hbar\Omega$ calculation using an phenomenological effective
interaction, but at an even lower energy of about 9 MeV
(see, however, Ref.~\cite{millener} for a comment on this work).

\subsection{$^6\mbox{Li}$} \label{Li6s}
For this nucleus, we perform a full $6\hbar\Omega$ calculation
($N_{\rm {max}} = 8$) for the positive-parity states
and a full $5\hbar\Omega$ calculation
($N_{\rm {max}} = 7$) for the negative-parity states.
The results are shown in Table IV and in Fig.4.
The six low-lying states known experimentally are nicely reproduced
except that the $J^{\pi}=2^+, \, T=1$ state at 5.37 MeV
and the $J^{\pi}=1^+\, T=1$ state at 5.65 MeV are obtained at excitation
energies about 1 MeV too high. The other four ``$0\hbar\Omega$''
states are obtained at excitation energies of 9.94,
10.74, 11.38 and 12.93 MeV.
The new results presented here again show some improvement
over the previous results \cite{lighta,mfd}. The excitation energies
for the first and the second excited states are closer to experiment
than those obtained in the GFMC calculations \cite{GFMC}.
In particular, the member of the $0^+$ isospin triplet state
is obtained at an excitation energy of 3.79 MeV, close to the
experimental value of 3.56 MeV. This state is of some interest
for the study of the isospin and parity violations \cite{li6}.

The lowest ``$2\hbar\Omega$'' state that we obtain
has $J^{\pi}=1^+, \, T=0$ and an excitation energy of 14.72 MeV.
It has the configuration of
$   2\% |0\hbar\Omega\rangle
+  56\% |2\hbar\Omega\rangle
+  22\% |4\hbar\Omega\rangle
+  19\% |6\hbar\Omega\rangle$.
We identify the second lowest ``$2\hbar\Omega$'' state at 16.08 MeV as the
experimental 15.8 MeV state \cite{li6_15.8}
since it has $J^{\pi}=3^+$.
This state has very little overlap with $0\hbar\Omega$ configurations;
its wave function can be expressed as
$  58\% |2\hbar\Omega\rangle
+  21\% |4\hbar\Omega\rangle
+  21\% |6\hbar\Omega\rangle$.

Below these ``$2\hbar\Omega$'' states we obtain five
negative-parity ``$1\hbar\Omega$'' states
with excitation energies of 10.9 to 14.2 MeV. One should not be surprised
if the experimental energies of these ``$1\hbar\Omega$'' states
turn out to be somewhat lower than the values listed in Table IV,
obtained in a $5\hbar\Omega$ ($N_{\rm {max}}=7$) calculation.
We have seen in the cases of $^4\mbox{He}$ and $^5\mbox{He}$ that
the excitation energies of the ``$1\hbar\Omega$'' states are lowered
by 1 to 3 MeV as we go from a $5\hbar\Omega$ space to a $7\hbar\Omega$ space.
However we are not able to perform a $7\hbar\Omega$ ($N_{\rm {max}}=9$)
calculation for the negative-parity states in
$^6\mbox{Li}$ at the present time.

A $6\hbar\Omega$ calculation for $^6\mbox{Li}$ was also attempted by
Bevelacqua \cite{beve} who used a modified Sussex interaction \cite{sussex}.
In that work, all experimentally known states were quite well reproduced.
But several ``$0\hbar\Omega$'' and ``$1\hbar\Omega$'' states that we obtain
here were not given in \cite{beve}. Some of these states, however, were
obtained by van Hees {\it et al.} \cite{vanH}
in a $(0+1)\hbar\Omega$ calculation using a phenomenological interaction.
For example, they obtained a $2^-$ state at an energy of about 9 MeV,
lower than our $2^-$ state at 10.86 MeV.

The g.s.~magnetic dipole moment is calculated to be $0.840\, \mu_N$,
slightly larger than the experimental value of $0.822\, \mu_N$.
The g.s.~quadrupole moment is calculated to be $-0.067\, e\,{\rm {fm}}^2$,
very close to the experimental value of $-0.082\, e\,{\rm {fm}}^2$.
These results are obtained by using bare electromagnetic operators.
In principle, these electromagnetic operators should also be
renormalized in a way consistent with how the effective interaction
is derived from the bare $NN$ potential.
This is particularly important when the model space
is small.  While we hope that our model spaces are large enough
to permit the use of bare operators, we are aware that this
assumption ought to be verified by explicit calculations
of effective operators.

\subsection{$^7\mbox{Li}$}
The negative-parity states are calculated in a
full $4\hbar\Omega$ space ($N_{\rm {max}}=7$)
and the positive-parity states in a full $5\hbar\Omega$ space
($N_{\rm {max}}=8$).
The results are given in Table V (see also Fig.5).
The theoretical spectrum appears expanded relative to experiment,
perhaps indicating that, for the model spaces we can handle, that the
two-body $G$ matrix is not an adequate approximation to $V^{\rm eff}$.

However, the energy of the first excited state ($1/2^-$)
agrees very well with experiment (0.46 MeV vs 0.48 MeV).
We had previously experienced some difficulty with this state in
single-$G$ calculations using smaller spaces \cite{lighta},
finding excitation energies that were too low.
The inclusion of high-lying unperturbed configurations is
important for reproducing this state at the experimental energy.
In a $0\hbar\Omega$ calculation, the excitation energy of this state
is only 0.195 MeV. When $2\hbar\Omega$ configurations are included,
the result increases to 0.498 MeV, which becomes 0.463 MeV when
$4\hbar\Omega$ configurations are taken into account.

The lowest positive-parity state we obtain has
$J^{\pi} = 1/2^+$ and $T=1/2$ and an excitation energy of
15.264 MeV. This state is dominated by the configurations
$(0s)^3 (0p)^4$ (about 50\%) $(0s)^2 (0p)^4 (1s)^1$ (12\%),
and $(0s)^4 (0p)^2 (1s)^1$ (10\%). The other 28\% is distributed
over many configurations.

For the g.s.~electric quadrupole moment $Q$, we obtain
$-2.37 \, e\, {\rm {fm}}^2$, much smaller in magnitude
than the experimental value of $-4.06 \, e\, {\rm {fm}}^2$.
We notice that the calculated quadrupole moment increases
in magnitude with the size of the model space. The results for
$Q$ from the $0\hbar\Omega$, $2\hbar\Omega$, and $4\hbar\Omega$
calculations are $-1.67$, $-2.16$ and $-2.37\, e\, {\rm {fm}}^2$,
respectively.  Presumably still larger model spaces are needed
to generate the degree of deformation indicated by the
quadrupole moment.
However, we are currently unable to go beyond $4\hbar\Omega$ for this nucleus.
One may notice from Table V that the calculated rms point charge radius is
also too small (2.06 fm versus 2.29 fm from experiment),
indicating the calculated wave function is probably confined to too small
a region by the limited size of the model space. We have repeated the
$4\hbar\Omega$ calculation for $\hbar\Omega = 11\, {\rm {MeV}}$.
The results are also listed in Table V. Although the rms radius
for this choice of $\hbar\Omega$ agrees quite well with experiment,
the result on the quadrupole moment $Q$ ($-2.67\, e {\rm {fm}}^2$),
though improved somewhat, is still too small.
Therefore changing the model space through adjustments in the basis parameter
$\hbar\Omega$ alone
is not sufficient, given our use of the bare operator;
one has to introduce higher configurations to realistically
describe this deformed nucleus.


\section{Dependence on the Size of the Model Space}
\label{size}

In this section we examine the differences arising from the use
of a multi-valued $G$ matrix, rather than
a conventional single-valued effective
interaction.  These differences are expected to diminish
as the model space is increased because the increasing
energy denominators in Eq.~(\ref{G}) suppress effects higher order in
$\tilde{V}$. In Table VI, the calculated energy and  root-mean-square (rms)
proton point radius of the ground state and the excitation energy
of the first excited state in $^4\mbox{He}$ are given for four different
model spaces ($N_{\rm {max}}=2, 4, 6$ and 8) and two choices of $\hbar\Omega$
(14 and 20 MeV). As expected (see also Fig.6),
the differences between the excitation energies obtained in
the conventional and multi-valued $G$-matrix calculations
diminish as the model spaces increase.
Similarly, the choice of $\hbar\Omega$ becomes less important in the
larger model spaces.
Note in particular that the calculated g.s.~rms radius is about the same
($\sim 1.49$ fm) in the $8\hbar\Omega$, multi-valued $G$ calculations for the
two values of $\hbar\Omega$, indicating good convergence for this
quantity.

It is clear from Table VI that the increased size of the
model space and the use of an appropriate (multi-valued) $G$ matrix both
contribute to the improved results for the $0^+_2$ state in $^4\mbox{He}$
in this work. For example, in a conventional (single-valued) $G$-matrix
calculation with $\hbar\Omega=14 \, {\rm {MeV}}$, the excitation energy of
this state decreases by 0.55 MeV from 22.93 MeV to 22.38 MeV when we go from
a $6\hbar\Omega$ space to a $8\hbar\Omega$ space;
in the $8\hbar\Omega$ space, the use of the multi-valued $G$ matrix
further decreases the result by another 0.56 MeV to 21.82 MeV.

As mentioned in Sec.~\ref{he4}, the relative importance of different
major-shell configurations depends on the choice of $\hbar\Omega$.
For $\hbar\Omega = 14$ MeV, the configurations of the $0^+_1$ and
$0^+_2$ states in $^4\mbox{He}$ obtained in $8\hbar\Omega$, multi-valued
$G$-matrix calculation are given in Eqs.~(\ref{gs14},\ref{es14}).
For $\hbar\Omega = 20$ MeV, we obtain
\begin{eqnarray}
[0^+_1] & \simeq & 86\% |0\hbar\Omega\rangle
                +  4\% |2\hbar\Omega\rangle
                +  5\% |4\hbar\Omega\rangle \nonumber \\
            & & +  2\% |6\hbar\Omega\rangle
                +  3\% |8\hbar\Omega\rangle    \label{gs20}
\end{eqnarray}
and
\begin{eqnarray}
[0^+_2] & \simeq & 0\% |0\hbar\Omega\rangle
                +  60\% |2\hbar\Omega\rangle
                +  20\% |4\hbar\Omega\rangle \nonumber \\
            & & +  15\% |6\hbar\Omega\rangle
                +   5\% |8\hbar\Omega\rangle.  \label{es20}
\end{eqnarray}
A comparison of the g.s.~configurations in Eqs.~(\ref{gs14}) and
(\ref{gs20}) for the two values of $\hbar\Omega$ shows that
$\hbar\Omega = 20$ MeV may be a more reasonable choice for the ground
state, since the g.s.~can be better approximated
as a $0\hbar\Omega$ state. However, from
Eqs.~(\ref{es14}) and (\ref{es20}), we can see that
the wave function of the first excited state
has stronger $6\hbar\Omega$ and $8\hbar\Omega$ components
(which means slower convergence with respect to the size of the model space)
for $\hbar\Omega = 20$ MeV than for $\hbar\Omega = 14$ MeV.
This is not surprising, as the $0^+_2$ state in $^4\mbox{He}$ is loosely
bound and has a much larger radius than the
$0^+_1$ state. Since it is generally much more difficult to obtain
a converged result for the $0^+_2$ state than for the $0^+_1$ state,
a basis which leads to faster convergence of the
$0^+_2$ state is obviously better when both states are desired.
In this sense, $\hbar\Omega = 14$ MeV is a better choice
than $\hbar\Omega = 20$ MeV for $^4\mbox{He}$.

\section{Conclusion}
\label{conclusion}
In a multi-$\hbar\Omega$ model space the two-body $G$ matrix is
dependent on the unperturbed energy of the other A-2 nucleons.
We have used such a multi-valued $G$ matrix
in large, no-core, shell-model calculations for light nuclei.
When compared to conventional calculations, proper treatment of the
$N^{\rm spectators}_{\rm sum}$ dependence of the $G$ matrix tends to
lower the energies of the ``$1\hbar\Omega$'' and ``$2\hbar\Omega$''
excited states more than the ``$0\hbar\Omega$'' states, bringing
energies into better agreement with experiment.

Applying this approach to large, no-core, shell-model calculations,
we have achieved a reasonable description of the ``low-lying''
states (including ``$1\hbar\Omega$'' and ``$2\hbar\Omega$'' states)
in light nuclei. With model spaces consisting of as many as nine
HO major shells, the experimentally known states in
$^4\mbox{He}$, $^5\mbox{He}$, $^6\mbox{Li}$, and $^7\mbox{Li}$
have been reproduced. Very good agreement with experiment has been
obtained for the excited states in $^4\mbox{He}$,
the ``single-particle'' $3/2^- -1/2^-$ splitting in $^5\mbox{He}$
and in $^7\mbox{Li}$, and the low-lying spectrum of $^6\mbox{Li}$,
etc.. Some earlier theoretical predictions of additional states
in the spectrum have been confirmed
[e.g., a $1/2^+$ state at 4.3 MeV and two nearly degenerate states
($3/2^+$ and $5/2^+$) at 9.7 MeV in $^5\mbox{He}$]. We have also
obtained a few low-lying states that have neither been
observed experimentally nor predicted theoretically before.
For example, we obtain a $3/2^-$ state at 12.0 MeV
in $^5\mbox{He}$ and several ``$0\hbar\Omega$'' and ``$1\hbar\Omega$''
states below 15 MeV in $^6\mbox{Li}$.
One shortcoming is that the calculated quadrupole moment of the
ground state in $^7\mbox{Li}$ is too small in magnitude when compared
with experiment. We attribute this disagreement to the relatively small
size ($4\hbar\Omega$) of the model space that is used for this nucleus;
it may be that the bare quadrupole operator is not appropriate for this
space.

By using large, no-core model spaces, we have eliminated
adjustable s.p.~energies usually involved with shell-model calculations
using effective interactions.
However, it should be emphasized that in calculating the $G$
matrices, we have used an empirical prescription for the starting energy,
which involves a parameter $\Delta$. This parameter is adjusted to yield
a reasonable binding energy. For this reason, our calculated
binding energies should not be confused as exact results, which can only
be obtained through a parameter-free approach. Recent
GFMC calculations of Pudliner {\it et al.} \cite{GFMC} serve as
a major step in this direction. Nevertheless,
we believe that once this parameter is adjusted
to reproduce the binding energy, other nuclear properties can then be
predicted.

There are important improvements that could be incorporated into
future calculations of the type reported here.  Our use of very
large model spaces was motivated by the hope that bare operators
and effective interactions approximated by a two-body $G$ matrix
might be successful in such spaces.  But presumably the need
for large values of $\Delta$ is connected with the omission of the folded
diagrams and neglected
interactions of three-body and higher clusters in the excluded space.
As there are prospects for improving these aspects of the
calculations $\cite{haxton}$, we consider the present
effort a first step toward the ultimate goal of accurate shell
model calculations based on realistic $NN$ interactions.

If one were able to generate the exact $V^{\rm eff}$,
energy eigenvalues should not depend on the choice of the model
space.  Thus perhaps the most important result from this initial exploration
of multi-valued $G$ matrices is that some improvement was achieved in the rate
of convergence of energy eigenvalues, as a function of the complexity
of the model space (see, for example, Fig.6). We would argue that
the degree to which our results can be further improved is an
open question: clearly we have the capacity to put substantial new
physics into calculations of $V^{\rm eff}$ and to
generate the corresponding effective operators.

\acknowledgements

We thank John Millener and Karlheinz Langanke for useful communications.
B.R.B. and D.C.Z. acknowledge
partial support of this work by the National Science Foundation,
Grant No.~PHY93-21668.
J.P.V. acknowledges partial support by the U.S.
Department of Energy under Grant No.~DE-FG02-87ER-40371, Division
of High Energy and Nuclear Physics.
W.C.H. and C.L.S. acknowledge
partial support of this work by the U.S.~Department of Energy.

\pagebreak

\begin{table}

TABLE I. Some diagonal matrix elements
$\langle (ab:JT) | V^{\rm eff} | (ab:JT) \rangle$
(in MeV) for four possible values of $N_{\rm sum}$ in a full $6\hbar\Omega$
calculation of the positive-parity states in $^6\mbox{Li}$.
\begin{tabular}{c|cccc}
         $N_{\rm sum}$               &    2   &   4    &   6    & 8\\ \hline
$(ab:JT) = (0s_{1/2}\,0s_{1/2}: 01)$ & -6.689 & -6.734 & -6.894 & -7.371 \\
$(ab:JT) = (0s_{1/2}\,0s_{1/2}: 10)$ & -8.272 & -9.006 & -9.969 &-11.554 \\
$(ab:JT) = (0s_{1/2}\,0p_{3/2}: 10)$ & -1.144 & -1.415 & -1.769 & -2.344 \\
$(ab:JT) = (0s_{1/2}\,0p_{3/2}: 11)$ & -3.768 & -3.812 & -3.935 & -4.273 \\
$(ab:JT) = (0s_{1/2}\,0p_{3/2}: 20)$ & -8.272 & -9.006 & -9.969 &-11.554 \\
$(ab:JT) = (0s_{1/2}\,0p_{3/2}: 21)$ & -1.006 & -1.029 & -1.058 & -1.090 \\
$(ab:JT) = (0p_{3/2}\,0p_{3/2}: 01)$ & -3.227 & -3.256 & -3.342 & -3.588 \\
$(ab:JT) = (0p_{3/2}\,0p_{3/2}: 10)$ & -1.272 & -1.575 & -1.950 & -2.522 \\
$(ab:JT) = (0p_{3/2}\,0p_{3/2}: 21)$ & -1.364 & -1.389 & -1.439 & -1.545 \\
$(ab:JT) = (0p_{3/2}\,0p_{3/2}: 30)$ & -4.179 & -4.528 & -5.021 & -5.821
\end{tabular}
\end{table}

\begin{table}

Table II. The results for ${}^4\mbox{He}$ from a full
$8\hbar\Omega$ [$N_{\rm {max}}=8$]
calculation for the positive-parity states and a full
$7\hbar\Omega$ [$N_{\rm {max}}=7$]
calculation for the negative-parity states.
In the Table, $E_B$ is the binding energy and $E_x(J_n^{\pi},T)$
the excitation energy of the $J^{\pi}_n,T$ state. All energies are
in MeV. The dominant major-shell configuration for each state is given
in the column labeled ``Main Conf.''.
The g.s.~rms {\em point} radius for protons
$\sqrt{\langle r^2_p \rangle}$ is also given.
The ``experimental'' g.s.~rms radius is deduced from the charge rms radius
$\sqrt{ \langle r^2_c \rangle }$ through (ignoring the neutron
charge distribution and other higher-order effects and
assuming a proton rms charge radius of 0.81 fm)
$\langle r^2_p \rangle = \langle r^2_c \rangle - 0.81^2$.
\begin{tabular}{c|ccc}
Observable & Main Conf. & Mult-valued $G$ & Experiment$^{a)}$ \\ \hline
$E_B$          &  ---            & 26.459  & 28.296\\
$\sqrt{\langle r^2_p\rangle}$ (fm)
               &  ---            &  1.492  & 1.46 \\
$E_x(0^+_1,0)$ & $0\hbar\Omega$  &    0    & 0 \\
$E_x(0^+_2,0)$ & $2\hbar\Omega$  & 21.824  & 20.21\\
$E_x(0^-_1,0)$ & $1\hbar\Omega$  & 21.566  & 21.01\\
$E_x(2^-_1,0)$ & $1\hbar\Omega$  & 23.003  & 21.84\\
$E_x(2^-_1,1)$ & $1\hbar\Omega$  & 24.214  & 23.33\\
$E_x(1^-_1,1)$ & $1\hbar\Omega$  & 24.418  & 23.64\\
$E_x(1^-_1,0)$ & $1\hbar\Omega$  & 25.286  & 24.25\\
$E_x(0^-_1,1)$ & $1\hbar\Omega$  & 25.370  & 25.28\\
$E_x(1^-_2,1)$ & $1\hbar\Omega$  & 25.671  & 25.95
\end{tabular}
\noindent
$^{a)}$ From Ref.~\cite{he4exp} except for the rms radius which is from
        Ref.~\cite{rms1}.
\end{table}

\begin{table}

Table III. The results for ${}^5\mbox{He}$ from a full
$6\hbar\Omega$ [$N_{\rm {max}}=7$]
calculation for the negative-parity states and a full
$7\hbar\Omega$ [$N_{\rm {max}}=8$]
calculation for the positive-parity states. Calculated states
with an excitation energy larger than 23 MeV are not shown.
All the states listed in this table have an isospin $T=1/2$.
The g.s.~electric quadrupole moment $Q$
and magnetic dipole moment $\mu$ are also listed.
See the caption of Table II for more explanations.
\begin{tabular}{c|ccc}
Observable  & Main Conf. & Multi-valued $G$ & Experiment$^{a)}$ \\ \hline
$E_B$                &  ---   & 25.883 & 27.410\\
$\sqrt{\langle r^2_p\rangle}$ (fm)
                     &  ---   &  1.630 & N/A \\
$\mu (\mu_N)$        &  ---   & -1.847 & N/A \\
$Q (e\, {\rm {fm}}^2)$ &  ---   & -0.443 & N/A \\
$E_x({3/2}^-_1)$ & $0\hbar\Omega$  &    0   & 0 \\
$E_x({1/2}^-_1)$ & $0\hbar\Omega$  &  2.465 & $4\pm 1^{b)}$\\
$E_x({1/2}^+_1)$ & $1\hbar\Omega$  &  4.343 & see $^{c)}$\\
$E_x({3/2}^+_1)$ & $1\hbar\Omega$  &  9.717 & see $^{c)}$\\
$E_x({5/2}^+_1)$ & $1\hbar\Omega$  &  9.727 & see $^{c)}$\\
$E_x({3/2}^-_2)$ & $2\hbar\Omega$  & 12.006 & N/A \\
$E_x({1/2}^-_2)$ & $2\hbar\Omega$  & 15.213 & N/A \\
$E_x({7/2}^-_1)$ & $2\hbar\Omega$  & 17.252 & N/A \\
$E_x({5/2}^-_1)$ & $2\hbar\Omega$  & 17.296 & N/A \\
$E_x({3/2}^+_2)$ & $1\hbar\Omega$  & 19.060 & 16.75 \\
$E_x({1/2}^+_2)$ & $1\hbar\Omega$  & 19.895 & N/A \\
$E_x({7/2}^+_1)$ & $1\hbar\Omega$  & 21.908 & N/A \\
$E_x({1/2}^+_3)$ & $1\hbar\Omega$  & 22.187 & N/A \\
$E_x({9/2}^+_1)$ & $1\hbar\Omega$  & 22.723 & N/A
\end{tabular}

\noindent
$^{a)}$ From Ref.~\cite{a5to10}.

\noindent
$^{b)}$ Analyses of experiments give different values ranging from
        1.4 MeV to 5.5 MeV.

\noindent
$^{c)}$ Previous theoretical works predict a $1/2^+$ state at
5-7 MeV and a $3/2^+$ and a $5/2^+$ state at 12-14 MeV,
see Refs.~\cite{hees,barker}.

\end{table}

\begin{table}

Table IV. The results for ${}^6\mbox{Li}$ from a full
$6\hbar\Omega$ [$N_{\rm {max}}=8$]
calculation for the positive-parity states and a full
$5\hbar\Omega$ [$N_{\rm {max}}=7$]
calculation for the negative-parity states.
Calculated states
with an excitation energy larger than 18 MeV are not shown.
See the caption of Table II for more explanations.
\begin{tabular}{c|ccc}
Observable  & Main Conf. & Multi-valued $G$ & Experiment$^{a)}$ \\ \hline
$E_B$                & ---             & 30.525 & 31.996\\
$\sqrt{\langle r^2_p\rangle}$ (fm)
		     & ---             &  2.11  &  2.41 \\
$\mu (\mu_N)$        & ---             &  0.840 &  0.822 \\
$Q (e\, {\rm {fm}}^2)$ & ---             & -0.067 & -0.082 \\
$E_x(1^+_1,0)$       & $0\hbar\Omega$  &  0     & 0   \\
$E_x(3^+_1,0)$       & $0\hbar\Omega$  &  2.619 & 2.186  \\
$E_x(0^+_1,1)$       & $0\hbar\Omega$  &  3.786 & 3.563  \\
$E_x(2^+_1,0)$       & $0\hbar\Omega$  &  4.713 & 4.31  \\
$E_x(2^+_1,1)$       & $0\hbar\Omega$  &  6.406 & 5.366 \\
$E_x(1^+_2,0)$       & $0\hbar\Omega$  &  6.764 & 5.65  \\
$E_x(2^+_2,1)$       & $0\hbar\Omega$  &  9.942 & N/A \\
$E_x(1^+_1,1)$       & $0\hbar\Omega$  & 10.742 & N/A \\
$E_x(2^-_1,0)$       & $1\hbar\Omega$  & 10.863 & N/A \\
$E_x(1^-_1,0)$       & $1\hbar\Omega$  & 11.082 & N/A \\
$E_x(1^+_3,0)$       & $0\hbar\Omega$  & 11.382 & N/A \\
$E_x(0^+_2,1)$       & $0\hbar\Omega$  & 12.934 & N/A \\
$E_x(0^-_1,0)$       & $1\hbar\Omega$  & 13.147 & N/A \\
$E_x(1^-_1,1)$       & $1\hbar\Omega$  & 13.706 & N/A \\
$E_x(2^-_1,1)$       & $1\hbar\Omega$  & 14.242 & N/A \\
$E_x(1^+_4,0)$       & $2\hbar\Omega$  & 14.716 & N/A \\
$E_x(1^-_2,0)$       & $1\hbar\Omega$  & 15.422 & N/A \\
$E_x(3^+_2,0)$       & $2\hbar\Omega$  & 16.083 & 15.8 \\
$E_x(2^-_2,0)$       & $1\hbar\Omega$  & 16.950 & N/A \\
$E_x(0^-_1,1)$       & $1\hbar\Omega$  & 17.328 & N/A \\
$E_x(0^+_3,1)$       & $2\hbar\Omega$  & 17.515 & N/A

\end{tabular}

\noindent
$^{a)}$ From Ref.~\cite{a5to10} except for the rms radius which is from
        Ref.~\cite{rms2}.

\end{table}

\begin{table}

Table V. The results for ${}^7\mbox{Li}$ from a full
$4\hbar\Omega$ [$N_{\rm {max}}=7$]
calculation for the negative-parity states and a full
$5\hbar\Omega$ [$N_{\rm {max}}=8$]
calculation for the positive-parity states.
Calculated states
with an excitation energy larger than 16 MeV are not shown.
See the caption of Table II for more explanations.
In this table, we also list the results for $\hbar\Omega = 11\, {\rm {MeV}}$.

\begin{tabular}{c|cccc}
Observable  & Main Conf. & $\hbar\Omega$=14 & $\hbar\Omega$=11
 & Experiment$^{a)}$ \\ \hline
$E_B$                & ---    & 37.533 & 36.141 & 39.244\\
$\sqrt{\langle r^2_p\rangle}$ (fm)
                     & ---    &  2.062 &  2.233  &  2.29 \\
$\mu (\mu_N)$        & ---    &  3.027 &  3.014  &  3.2564 \\
$Q (e\, {\rm {fm}}^2)$ & ---    & -2.372 & -2.672  & -4.06 \\
$E_x({3/2}^-_1,1/2)$ & $0\hbar\Omega$  &    0   &  0     & 0     \\
$E_x({1/2}^-_1,1/2)$ & $0\hbar\Omega$  &  0.463 &  0.188 & 0.4776\\
$E_x({7/2}^-_1,1/2)$ & $0\hbar\Omega$  &  5.249 &  5.872 & 4.63  \\
$E_x({5/2}^-_1,1/2)$ & $0\hbar\Omega$  &  7.325 &  7.006 & 6.68  \\
$E_x({5/2}^-_2,1/2)$ & $0\hbar\Omega$  &  8.857 &  8.473 & 7.4595\\
$E_x({3/2}^-_2,1/2)$ & $0\hbar\Omega$  & 10.749 &  9.178 & 9.85  \\
$E_x({7/2}^-_2,1/2)$ & $0\hbar\Omega$  & 11.402 & 11.012 & 9.67  \\
$E_x({1/2}^-_2,1/2)$ & $0\hbar\Omega$  & 11.608 & 10.165 & N/A   \\
$E_x({5/2}^-_2,1/2)$ & $0\hbar\Omega$  & 12.847 & 11.920 & N/A   \\
$E_x({3/2}^-_1,3/2)$ & $0\hbar\Omega$  & 12.961 & 12.592 & 11.24 \\
$E_x({3/2}^-_3,1/2)$ & $0\hbar\Omega$  & 13.237 & 11.837 & N/A   \\
$E_x({1/2}^-_3,1/2)$ & $0\hbar\Omega$  & 13.704 & 12.419 & N/A   \\
$E_x({1/2}^+_1,1/2)$ & $1\hbar\Omega$  & 15.264 &        & N/A   \\
$E_x({1/2}^-_1,3/2)$ & $0\hbar\Omega$  & 15.780 &        & N/A
\end{tabular}

\noindent
$^{a)}$ From Ref.~\cite{a5to10} except for the rms radius which is from
        Ref.~\cite{rms2}.

\end{table}

\begin{table}

Table VI. The results for the g.s.~energy (in MeV),
proton rms radius (in fm) and the excitation energy (in MeV) of the
first excited state in ${}^4\mbox{He}$ obtained in the
multi-valued $G$ (m-$G$) and single-$G$ (s-$G$) calculations in
different model spaces with two choices of $\hbar\Omega$ (14 and 20 MeV).
The difference between the s-$G$ and m-$G$ results is also given.
\begin{tabular}{ccc|ccc}
$\hbar\Omega$ & $N_{\rm {max}}$ & Approach &
    $E(0^+_1)$ & $\sqrt{\langle r^2_p\rangle}$ & $E_x(0^+_2)$ \\  \hline
14 & 2 & s-$G$  &-23.18 & 1.57 & 26.38 \\
   &   & m-$G$  &-23.64 & 1.56 & 25.17 \\
   &   & diff.  &  0.46 & 0.01 &  1.21 \\ \cline{2-6}
   & 4 & s-$G$  &-25.23 & 1.57 & 26.73 \\
   &   & m-$G$  &-25.95 & 1.56 & 25.78 \\
   &   & diff.  &  0.72 & 0.01 &  0.95 \\  \cline{2-6}
   & 6 & s-$G$  &-25.62 & 1.51 & 22.93 \\
   &   & m-$G$  &-26.44 & 1.49 & 22.27 \\
   &   & diff.  &  0.82 & 0.02 &  0.66 \\  \cline{2-6}
   & 8 & s-$G$  &-25.62 & 1.51 & 22.38 \\
   &   & m-$G$  &-26.46 & 1.49 & 21.82 \\
   &   & diff.  &  0.84 & 0.02 &  0.56 \\ \hline
20 & 2 & s-$G$  &-25.62 & 1.38 & 33.05 \\
   &   & m-$G$  &-25.94 & 1.37 & 30.56 \\
   &   & diff.  &  0.32 & 0.01 &  2.49 \\  \cline{2-6}
   & 4 & s-$G$  &-26.34 & 1.46 & 31.84 \\
   &   & m-$G$  &-26.84 & 1.45 & 30.23 \\
   &   & diff.  &  0.50 & 0.01 &  1.61 \\  \cline{2-6}
   & 6 & s-$G$  &-25.73 & 1.46 & 26.93 \\
   &   & m-$G$  &-26.27 & 1.46 & 25.49 \\
   &   & diff.  &  0.54 & 0.00 &  1.44 \\  \cline{2-6}
   & 8 & s-$G$  &-25.21 & 1.49 & 24.71 \\
   &   & m-$G$  &-25.82 & 1.48 & 23.35 \\
   &   & diff.  &  0.61 & 0.01 &  1.36 \\ \hline
\multicolumn{3}{c|}{Experiment}
                   &-28.30 & 1.46 & 20.21
\end{tabular}
\end{table}

\section*{Figure Captions}

\subsection*{Figure 1}
An illustration of the $Q$ operator appropriate for
a full $6\hbar\Omega$ calculation of $^6\mbox{Li}$.
The regions interior to the lines are the $Q$=0 regions defined
in Eq.~(\ref{Q0A}). The lines correspond to the possible values
of $N_{\rm sum}^{\rm spectators}$, which range from 0 to $N_{\rm max}$.
The contour for $N_{\rm sum}^{\rm spectators} = 0$
is given as a solid line.  The wings result from the fact that the
spectator nucleons are in a unique configuration (closed $0s$ shell),
in this case, forbidding scattering into the $0s$ shell.
The wings make a negligible contribution numerically and can be ignored.
The contours for other values of
$N_{\rm sum}^{\rm spectators}$ are denoted by dashed lines.
Note that a single-valued $G$ matrix would employ
a single contour and thus neglect much of the physics governing
$V^{\rm eff}$ in a multi-$\hbar\Omega$ space.

\subsection*{Figure 2}
The calculated and experimental low-lying energy spectrum of
$^4\mbox{He}$.

\subsection*{Figure 3}
The calculated and experimental low-lying energy spectrum of
$^5\mbox{He}$. The first excited state ($1/2^-$) is very
broad; its experimental excitation energy is not well defined.
Refs.~\cite{hees,barker} also predict a low-lying
$1/2^+$ state at 5-7 MeV and $3/2^+$ and $5/2^+$ states at 12-14 MeV.
We obtain a few ``$2\hbar\Omega$'' (relative to the g.s.)
states (e.g., a $3/2^-$ state at 12.88 MeV) that have not been
observed experimentally nor predicted theoretically before.
All the states shown in this Figure have an isospin $T=1/2$.

\subsection*{Figure 4}
The calculated and experimental low-lying energy spectrum of
$^6\mbox{Li}$.

\subsection*{Figure 5}
The calculated and experimental low-lying energy spectrum of
$^7\mbox{Li}$. All the states shown in this figure
have an isospin $T=1/2$ except
for the $3/2^-$ state at 12.96 MeV which has $T=3/2$.

\subsection*{Figure 6}

(a) The rms point charge radius of the ground state
in $^4\mbox{He}$ obtained in multi-valued $G$ (mG)
calculations with model spaces of different sizes
[signified by $N_{\rm {max}}$]
using two values of the HO basis parameter $\hbar\Omega$
[14 MeV (solid lines) and 20 MeV (dashed lines)].

(b) Similar to (a) but for the excitation energy of the first excited $0^+$
state in $^4\mbox{He}$. Results from both the multi-valued $G$ (mG)
and single-$G$ (sG) calculations are shown.

\end{document}